%% file: hldvt2012.tex
\documentclass[10pt,journal,letterpaper]{IEEEtran}

\usepackage{makeidx}  
\usepackage[noend,ruled,linesnumbered]{algorithm2e}
\usepackage{lscape}
\usepackage{rotating}
\usepackage{verbatim}
\usepackage{symbols}
\usepackage{graphicx}
\usepackage{color}
\usepackage[cmex10]{amsmath}
\usepackage{amsfonts}
\begin{document}

\newtheorem{Definition}{Definition}[section]

\title{Using Decision Diagrams to Compactly Represent the State Space for Explicit Model Checking}
\author{Hao Zheng, Andrew Price, and Chris Myers \\
Computer Science and Engineering, University of South Florida, Tampa, FL 33620
\thanks{This material is based upon work supported by the National Science Foundation under Grant No. 0546492 and 0930510.
Any opinions, findings, and conclusions or recommendations expressed in this material are those of the author(s) and do not necessarily reflect the views of the National Science Foundation.}
\thanks{Hao Zheng and Andrew Price are with the Dept. of Computer Science and Engineering at the University of South Florida, Tampa, FL. Chris Myers is with the Dept. of Electrical and Computer Engineering at the University of Utah, SLC, UT.}
}

\pagestyle{empty}
\thispagestyle{empty}
\maketitle

\begin{abstract}
The enormous number of states reachable during explicit model checking is the main bottleneck for scalability.  This paper presents approaches of using decision diagrams to represent very large state space compactly and efficiently.  This is possible for asynchronous systems as two system states connected by a transition often share many same local portions.  Using decision diagrams can significantly reduce memory demand by not using memory to store the redundant information among different states.   This paper considers multi-value decision diagrams for this purpose.  Additionally, a technique to reduce the runtime overhead of using these diagrams is also described.  Experimental results and comparison with the state compression method as implemented in the model checker SPIN show that the approaches presented in this paper are memory efficient for storing large state space with acceptable runtime overhead.   
\end{abstract}


\IEEEpeerreviewmaketitle

\section{Introduction}
\label{intro}

Model checking \cite{Clarke00} is an automated formal analysis method for verifying hardware and software systems. It systematically checks whether a model of a given system satisfies a desired property such as deadlock freedom and request-response properties \cite{BAIER08}.  However, model checking is computationally very expensive as it searches the entire reachable state space of a design for errors.  Typically, all reachable states found during reachability analysis need to be stored to avoid searching the same part of state space multiple times unnecessarily.  As the number of states grows exponentially in the size of designs under verification, physical memory installed on typical computers can be exhausted quickly, therefore limiting model checking to the designs of small sizes.  This problem is well known as state explosion.   

In asynchronous designs, multiple components execute concurrently.  When verifying asynchronous designs, generally interleaving semantics are used to represent the behavior of such designs.  More specifically, when multiple components are ready to execute in a state, only one component is selected.  This interleaved execution makes sure all possible orderings of concurrently enabled executions are considered, thus capturing all possible behavior for verification.  The need to consider all possible interleavings of concurrent executions is the main cause of state explosion in asynchronous design verification as the number of interleavings grows exponentially if a design has a high degree of concurrency, and this leads to an excessively large state space for even a relatively small system.

In the interleaved execution as described above, only one component executes and updates the relative portion of a state, and the remaining portions of the state are unchanged.  Therefore, storing the entire states results in unnecessary overhead.   In model checker SPIN \cite{holzmann:97}, a state compression method is described where the independent parts of the design states are considered when representing states.  Independent parts refer to variables including global and local variables belonging to different processes.  An unique copy is created for each different  independent part, and the references to the copies of independent parts are used to construct the states. 
   
While the above compression method reduces memory significantly in representing states, there are still a lot of redundancies in the state representations.  As pointed out in the previous paragraphs, a global state of a design is in fact a tuple of pointers to the local states.  Two states connected by a state transition may still share a large number of same pointers in their representations as a state transition typically causes only a few local changes.  Therefore, reducing the redundancy in the state representations further can improve the scalability even more.  This paper proposes to use the multi-value decision diagrams to represent the reachable states for explicit model checking of asynchronous systems.  The multi-value decision diagrams, similar to the well-known binary decision diagrams (BDD) \cite{bryant92:bdd}, uses directed acyclic graphs to sets of objects encoded by typed variables.  The compactness of the decision diagram representations is due to sharing of common portions of a large number of different objects.    

BDDs are widely used in symbolic model checking \cite{Burch:SMC} where the model checking algorithms and the state representations are both based on the Boolean operations.  However, in the explicit model checking, individual states are created and manipulated.  Adding individual states into a decision diagram representation can incur high runtime overhead, and this would cancel largely the benefit of compact memory footprint of the decision diagrams.  To address this problem, this paper also proposes to use multi-value decision trees along with the decision diagrams.  Adding states into a decision tree is much faster, but it requires more memory than a decision diagram.  In this method, a decision tree is used as a buffer where it stores the reachable states until some threshold  is exceeded.  Then, the decision tree is compressed, and merged with the decision diagram representing the whole reachable state space.  This approach can significantly reduce the runtime overhead associated with using the decision diagrams, even though it may increase the average memory required compared to using the decision diagrams solely.   

In this paper, the multi-value decision diagrams are only used to store the states found during the depth-first search  that are generally implemented for the explicit model checking of asynchronous systems.  The algorithms themselves remain the same.  Therefore, the previous results to improve scalability such as partial order reduction \cite{Holzmann94,Patrice95} can still be used.   

The idea of using graph representation to store reachable states is previously described in \cite{spin:book} where a finite automaton is used.  As indicated in \cite{spin:book}, this approach has very high overhead to use.  This paper addresses that problem by using a decision tree as a buffer to reduce such overhead.  Using multi-value decision diagrams for storing states is also described in \cite{Ciardo01saturation}.  There are two main differences between this work and that in \cite{Ciardo01saturation}.  First, the Petri-net models used in \cite{Ciardo01saturation} are different from the model used in this work where in this paper Petri-net transitions are labeled with additional state space information.  Second, \cite{Ciardo01saturation} uses an exotic search strategy to achieve high memory efficiency.  However, this work uses multi-value decision diagrams to store reachable states without altering the depth-first search framework, therefore partial order reduction can be easily integrated. 

\section{Background}
\label{background}

\subsection{Labeled Petri-Nets}
\label{LPN}

This paper uses \emph{Labeled Petri-Nets} to model asynchronous systems.
Petri-Nets are a common modeling formalism for asynchronous designs \cite{Murata89,Jared07}. A Petri-net is a directed graph with a set of transitions and a set of places. A labeled Petri-Net is a Petri-net where transitions are labeled with various information representing a system's properties and behavior \cite{Thacker10}. Its definition is given as follows.
	\begin{Definition}
	\label{LPN-def}
 				A labeled Petri-net (LPN) is a tuple $N = \langle V,P,T,F, \marking_0,\stvec_0,L \rangle$,where
        \begin{enumerate}
        \item $V$ is a set of state variables of the integer type,
        \item $P$ is a finite set of places,
        \item $T$ is a finite set of transitions,
        \item $F \subseteq (P \times T) \cup (T \times P)$ is a finite set of the flow relations,
        \item $\marking_0 \subseteq P$ is a finite set of initially marked places,
        \item $\stvec_0: V \rightarrow \mathbb{Z}$ is a labeling function that assigns each variable an initial value,
        \item $L = \langle Guard, Assign \rangle$ is a pair of labeling functions for transitions in $T$, which is defined below.
        \end{enumerate}
	\end{Definition}	
	
	\begin{figure*}[htb]
	\begin{center}
	\begin{tabular}{cc}
	\includegraphics[width=36mm]{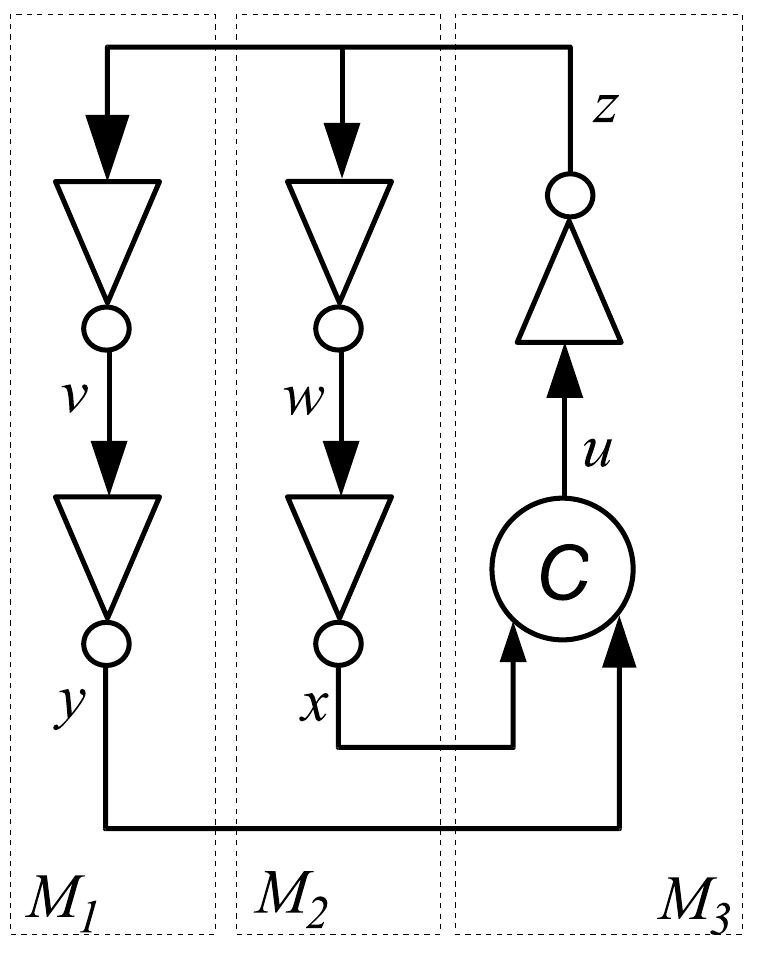} &
	\includegraphics[width=8cm]{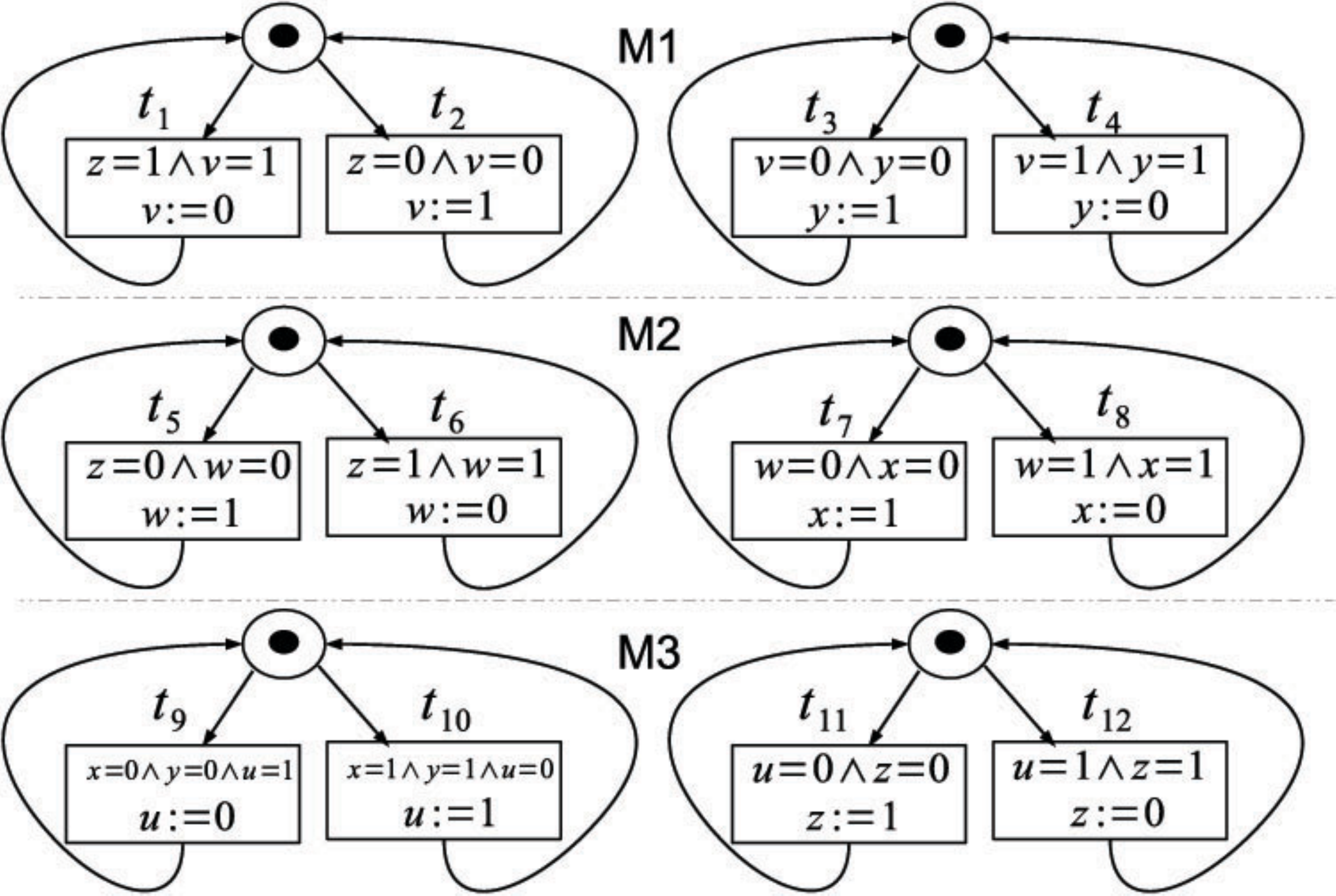} \\
	(a) &  (b)\\
	\end{tabular}
	{\caption{	\label{circuitLPN} (a) A simple asynchronous circuit. (b) The LPNs for module $M_1$, $M_2$, and $M_3$. The initial values of variables $u$, $v$, $w$, $x$, $y$, and $z$ are $0$, $1$, $1$, $0$, $0$, and $0$, respectively.}}
	\end{center}
	\end{figure*}

A simple LPN example is shown in Fig~\ref{circuitLPN}. Fig.\ref{circuitLPN}(a) shows a simple asynchronous circuit consisting of three components, and Fig.~\ref{circuitLPN}(b) shows the LPNs for each component in the circuit. For each component, its LPN has 2 places and 4 transitions.  The places are represented as circles, and the transitions are represented as boxes.
Each place is preceded and followed by one or more transitions, and each transition is preceded and followed by one or more places. The flow relations are represented by the edges connecting the transitions and places \cite{Jared07}.
The bullets found in some places are called tokens. Each place can have at most one token.  A place is {\em marked} if it has a token. A marking of LPN, $\marking \subseteq P$ is a set of marked places.

The dynamic behavior of a concurrent system is captured by LPN transitions with labelings.
Each transition $t \in T$ has a $preset$ denoted by $\bullet t = \{p \in P | (p,t) \in F \}$, which is the set of places connected to $t$, and a $postset$ denoted by $ t \bullet = \{p \in P | (t,p) \in F \}$, which is the set of places to which $t$ is connected. The $preset$ and $postset$ for places are defined similarly.

Before defining the transition labels formally, the grammar used by these labels is introduced first below \cite{Thacker10}.  The numerical portion of the grammar is defined as follows:
	\begin{eqnarray*}
  	\chi & ::= & c_i~|~v_i~|~(\chi)~|~-\chi~|~\chi+\chi~|~\chi-\chi~|~\chi*\chi~|\\
	 	 &     & \chi/\chi~|~\chi\verb!^!\chi~|~\chi\%\chi~|~\textup{NOT}(\chi)~|~\textup{OR}(\chi,\chi)~|\\ 	 
		 &     & \textup{AND}(\chi,\chi)~|~\textup{XOR}(\chi,\chi)~|~\textup{INT}(\phi)
	\end{eqnarray*}
where $c_i$ is an integer constant from ${\mathbb Z}$, and $v_i$ is an integer variable.
The functions NOT, OR, AND, and XOR are bit-wise logical operations assuming a 2's complement format with arbitrary precision.  $\textup{INT}(\phi)$ returns $1$ if the Boolean expression $\phi$, which is defined below, evaluates to \textbf{true}, or $0$ otherwise.
The set $\mathcal{P_\chi}$ is defined to be all formulas that can be constructed from the $\chi$ grammar.

The Boolean portion of the grammar is as follows:
	\begin{eqnarray*}
  	\phi & ::= & \textbf{true}~|~\textbf{false}~|~v_i~|~\neg \phi~|~\phi
  	\wedge \phi~|~\phi \vee \phi~|~\chi \equiv \chi~|\\ & & \chi\geq\chi~|~\chi>\chi~|~\chi\leq\chi~|~\chi<\chi
	\end{eqnarray*}
where the integer $v_i$ is regarded as \textbf{true} if its value is nonzero, and \textbf{false} otherwise.  In this sense, it is similar to the semantics of the C language.  The set $\mathcal{P_\phi}$ is defined to be all formulas that can be constructed from the $\phi$ grammar.

As in Definition~\ref{LPN-def}, each LPN transition is labeled with an enabling condition and a set of variable assignments.  LPN transition labeling is defined by $L = \langle Guard, Assign \rangle$ where
	\begin{itemize}
	\item $Guard : T \rightarrow \mathcal{P_\phi}$ labels each LPN transition with a Boolean expression that defines its enabling condition.
	\item $Assign: T \times V \rightarrow \mathcal{P_\chi}$ labels each LPN transition $t \in T$ and each variable $v \in V$ with an integer assignment made to $v$ when $t$ fires.
	\end{itemize}

For a large complex design, it usually consists of multiple components.  Then, each component is modeled in a LPN module, and the LPN for the whole design is the parallel composition of the LPN modules.  Let $N_1$, $\ldots$, $N_n$ be the LPN modules for the components consisting of a design.  The LPN for the design is $N = N_1 \| \ldots \| N_n$ where $\|$ is the parallel composition operator for LPNs.   In our method, communications among components are represented by the shared variables.  Shared variables between components $N_i$ and $N_j$ are variables in $V_i \cap V_j$.  Moreover, no two LPN modules share any common places.  In other words, $\forall_{i, j} i\not=j \imply P_i \cap P_j = \emptyset$.  In Fig.~\ref{circuitLPN}, the design is partitioned into three components for illustration purpose.  These components are represented by LPN modules as shown in $M_1$, $M_2$, and $M_3$.

\subsection{Reachability Analysis}
\label{reachability}

A basic approach for analyzing the dynamic behavior of a concurrent system modeled with LPNs is reachability analysis, which finds all possible state transitions and thus reachable states for such a system.  The reachable state space is typically represented by a state graph. A state graph is a directed graph where vertices represent states and edges represent state transitions.

A state of a LPN module $N_i$ is a pair $(\marking_i,\sigma_i)$ where $\marking_i$ denotes a marking of $N_i$ and $\sigma_i$ denotes a vector of values over variables $\varset_i$ of $N_i$.  The initial state $\initst$ of $N_i$ is $(\marking_0, \stvec_0)$.  Given a state $s_i$ of module $N_i$, $M(s_i)$ denotes the marking of $s_i$ and $\sigma(s_i)$ denotes the state vector of $s_i$.  Also, for any expression $e \in \mathcal{P_\chi} \cup \mathcal{P_\phi}$, $value(e, s_i)$ denotes a function that returns the value of expression $e$ in state $s_i$.  As described in the previous section, the LPN of a design is the parallel composition of a set of LPN modules, each of which represents a component in the design.  Let $N = N_1 \| \ldots \| N_n$ be the LPN of a design where $N_i~(1 \leq i \leq n)$ is the LPN module for the $i$th component of the design.  A global state $\gSt$ of $N$ is a $n$-tuple $(\lSt_1, \ldots, \lSt_n)$ where each $\lSt_i~(1 \leq i \leq n)$ is a local state of LPN module $N_i$.

Before describing the reachability analysis, the enabling condition of the LPN transitions is defined below.

	\begin{Definition}
	\label{enableTran-def}
 		Let $N_i$ be a LPN module.  A LPN transition $t$ is enabled in a state $s$ if the following two conditions are met:
        \begin{enumerate}
        \item $\bullet t \subseteq \marking(s)$,
        \item $value(e, s)$ is $true$ or not zero for $e = Guard(t)$.
        \end{enumerate}
	\end{Definition}

In Fig~\ref{circuitLPN}, every transition has its preset included in the initial marking.  In the initial state, the values of variable $u$ and $z$ are $0$, $Guard(t_{11})$, which is $u = 0 \wedge z = 0$, is evaluated to be true, therefore transition $t_{11}$ is enabled in the initial state.

Given a LPN module, the set of transitions enabled in a state $s$ is denoted by $enabled(s)$.  Naturally, the enabled transitions of the whole design LPN is denoted by $enabled(\gSt)$. The reachable state space of a LPN model can be found by exhaustively firing every enabled transition starting at the initial state. Firing a transition may lead to a new state by generating a new marking and a new state vector according to the assignments labeled for such transition.  Detailed definition of transition firing can be found in \cite{Jared07}.  In this paper, $s^\prime = t(s)$ denotes that a new state $s^\prime$ is produced by firing transition $t$ in state $s$ in a LPN module.  Similarly, $\gSt^\prime = t(\gSt)$ denotes that a new global state $\gSt^\prime$ is produced by firing transition $t$ in global state $\gSt$.



\begin{algorithm}
\caption{\label{basicSearch}$search((T,P,F,M_0))$}
\BlankLine
$stateTable.add(\gSt_0)$\;
$stateStack.push(\gSt_0)$\;
$enabledStack.push(enabled(\gSt_0))$\;
\While{stack is not empty}{
		$\gSt = stateStack.top()$\;
		$E_{\gSt} = enabledStack.top()$\;
		\If{$E_{\gSt} = \emptyset$} {
			$stateStack.pop()$\;
			$enabledStack.pop()$\;
			{\bf continue}\;
		}
		Select $t \in E_{\gSt}$ to fire\;
		$E_{\gSt} = E_{\gSt} \backslash t$\;
		$\gSt^\prime = t(\gSt)$\;
		\If{$stateTable.contains(\gSt)=={\bf true}$}{
			$stateStack.push(\gSt^\prime)$\;
			$enabledStack.push(enabled(\gSt^\prime))$\;
			$stateTable.add(\gSt^\prime)$\;		
		}
}
\end{algorithm}

The procedure to find the reachable state space of a given LPN model is given in Algorithm~\ref{basicSearch}.  Reachable states are stored in \emph{stateTable}.  During the reachability analysis, after a LPN transition $t_i$ is fired in global state $\gSt$, the local state of $M_i$ is changed to a new one.  In other words, if transition $t_i$ is fired in state $\gSt = (s_1, \ldots, \lSt_i, \ldots, s_n)$, a new global state $\gSt^\prime$ is generated such that $\gSt^\prime = (\lSt^\prime_1, \ldots, \lSt^\prime_i, \ldots, \lSt^\prime_n)$.  Observe that in $\gSt$ and $\gSt^\prime$, many local states might be exactly the same.  To avoid creating fresh copies of the same local states for $\gSt^\prime$ in memory, the actual definitions of the distinct local states are stored in a hash table for each LPN module, and the pointers to the local states are used to construct the global states.  From now on, $\gSt = (s_1, \ldots, s_n)$ is viewed as a tuple of pointers to the local states $s_i, \ldots$, and $s_n$.  In this way, a local state is created only once, and its pointer may be referenced many times in different global states. The similar state representation is also implemented in the model checker SPIN \cite{holzmann:97} and Java PathFinder \cite{Brat00}.

In many existing model checkers, the obvious choice of the data structure for $stateTable$ is hash tables.  Even though inserting and accessing hash tables are usually very efficient, the memory usage increases exponentially as the number of states found during the reachability analysis increases exponentially in the size of a design.  Therefore, representing and storing reachable states compactly is extremely important.   To address this problem, the following sections describe how the reachable states are represented using graphs, specifically decision trees and diagrams which may allow very large number of states to be stored in a relatively small memory footprint.

\section{State Space Representations}
\label{StateRep}

As described above, different global states might have a large number of pointers to the same local states.  To save memory even further by avoiding storing the same local state pointers in different global states, a data structure called {\em multi-value} decision diagram (MDD) is implemented in our method to store global states found during the reachability analysis.  The concept of MDD \cite{srinivasan1990} is very close to that of binary decision diagrams (BDDs) \cite{bryant92:bdd} in that both use graphs to represent a set of objects encoded by some variables.  The memory efficiency comes from the sharing of the same variables used to encode different objects.  In BDDs, the variables used for the encoding are binary variables, while integer variables are used for the encoding in the case of MDDs.  In this sense, BDDs can be regarded as a special case of MDDs.
The structural representation of the global states in our method is natural for MDDs. To facilitate the presentation, an unique integer index is assigned to each local state of a single LPN module.  Therefore, a global state $\gSt = (s_1, \ldots, s_n)$ can be regarded as a tuple of integers where $s_i$ is the index to a local state of module $i$.

\subsection{Multi-Value Decision Trees}
\label{MDT}

\begin{figure}
\begin{center}
\begin{tabular}{cc}
\begin{minipage}{1in}
\begin{eqnarray*}
(0,0,0)\\
(0, 0, 1) \\
(0,1,0)\\
(0,1,1)\\
(1,1,1)\\
(1,1,2)\\
\end{eqnarray*}
\end{minipage}
&
\begin{minipage}{1.5in}
\begin{center}
\includegraphics[width=1.8in]{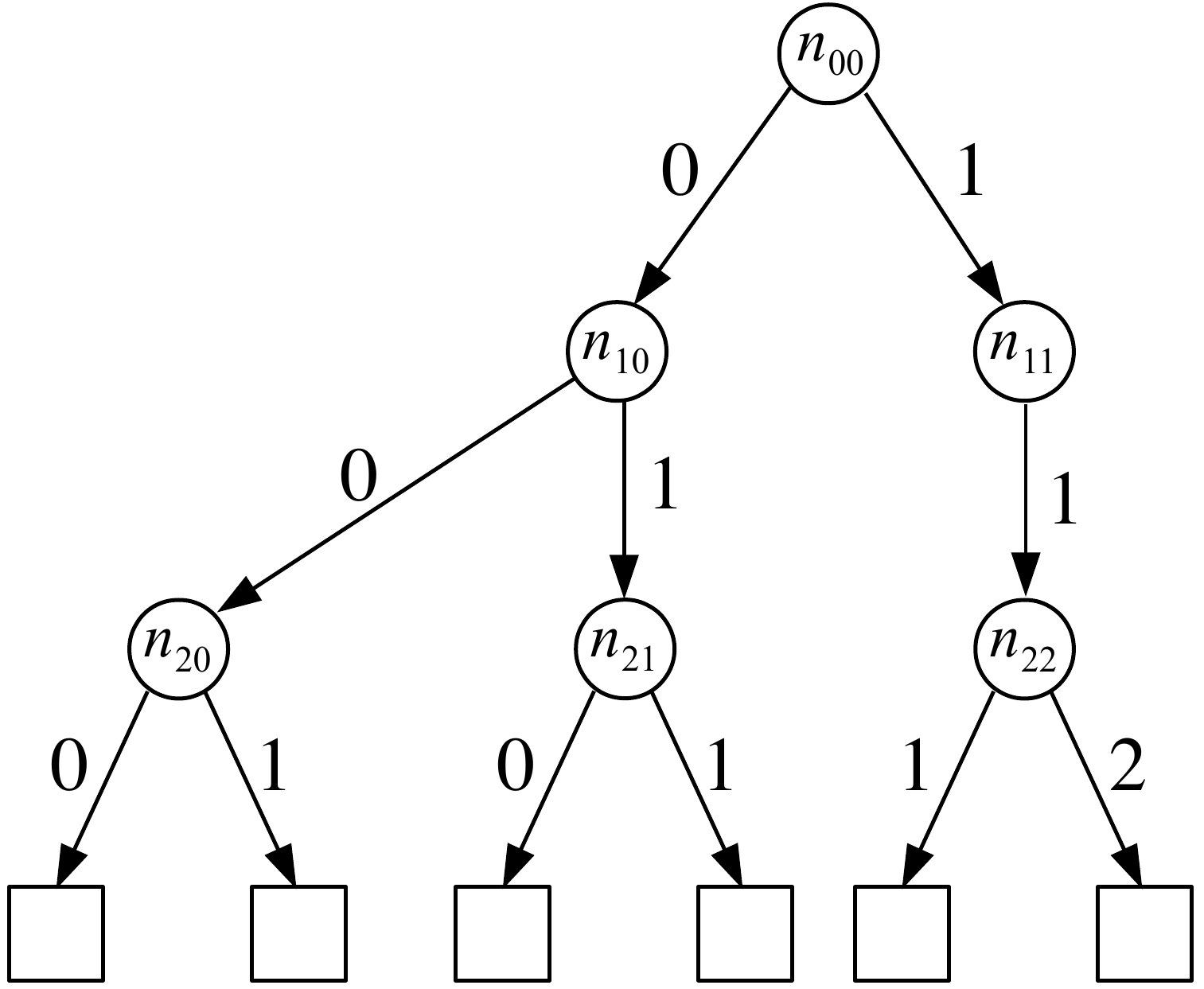}
\end{center}
\end{minipage} \\
(a) & (b)
\end{tabular}
\end{center}
\caption{\label{mdtree-ex1} (a) A set of integer triples, (b) The corresponding multi-value decision tree representation for the set shown in (a).}
\end{figure}

This section describes multi-value decision trees to introduce multi-value decision diagrams that are described in the next section.  A multi-value decision tree is a directed acyclic graph with a single root node, terminal nodes and non-terminal nodes.  The root node does not have incoming edges, and the terminal nodes do not have any outgoing edges.  Each non-terminal node has a single incoming edge and one or more outgoing edges.  Each edge is labeled with an integer number.  A path is a sequence of edges from the root to one of the terminal nodes, therefore an integer tuple can be formed by collecting the integers labeled on the edges on the path.  In other words, a path represents or encodes an integer tuple.  An example of multi-value decision tree is shown in Fig.~\ref{mdtree-ex1}.  Note that a terminal node is drawn for each path in the figure to make presentation clear.  In the actual implementation, only a single terminal node is used, and all paths converge to this unique terminal node. 

In our implementation, two operations are supported for multi-value decision tree, \texttt{add} and \texttt{contains}.  Function \texttt{add} takes as input an integer tuple, and creates necessary edges in the decision tree to build a path corresponding to the input tuple.  For example shown in Fig.~\ref{mdtree-ex1}, if a new tuple $(1,1,3)$ is given, it can be added to the decision tree by creating an edge from node $n_{21}$ to the terminal node, and this edge is labeled with $3$.  If another tuple $(1, 1, 1)$ is given, then no new edge is created as there is a path existing in the decision tree corresponding to the given tuple.  Function \texttt{contains} also takes as input an integer tuple, and checks if there is a path in the decision tree that corresponds to the given tuple.  It returns true if there is, false otherwise.  The time complexity of both functions is linear in the size of the tuples.  

The memory efficiency of the decision trees depends on the similarity of the prefixes of the tuples in the given set.  If there are a large number of tuples with long same prefixes, a lot of node sharing can reduce the memory usage significantly.  On the other hand, the worst case happens when a large number of tuples differ on their first elements.  In this case, more memory may be required compared to using hash tables as an edge is needed for each element in each tuple.    

\subsection{Multi-Value Decision Diagrams}
\label{MDD}

A multi-value decision diagram, similar to the decision trees, is a rooted directed acyclic graph where there is an unique root node, an unique terminal node, and non-terminal nodes.  In a multi-value decision diagram, the nodes are partitioned into levels.  The number of levels of a decision diagram is equal to the size of the integer tuples representing the global states. The root node is at level~0, which is at the top of the decision diagram. The node at the bottom is the terminal.  Each non-terminal node has a number of outgoing edges connecting nodes at one level higher or the terminal, and a number of incoming edges from nodes including the root node from one level lower.  Non-terminal nodes in the decision diagrams, unlike those in the decision trees, can have multiple incoming and outgoing edges.  A path from the root node to the terminal node in a decision diagram corresponds an integer tuple.   In the reachability analysis, the decision diagrams are used to replace the hash tables, therefore our implementation only supports two operations, \texttt{union} and \texttt{contains}.  Since \texttt{contains} for the decision diagrams is the same as function \texttt{contains} for the decision trees,  \texttt{union} is described in this section.  

Function \texttt{union} takes as inputs two decision diagrams, and returns a decision diagram that includes all the paths from either of the two input decision diagrams.  During the \texttt{union} operation, an unique table is maintained to make sure that no equivalent nodes are created.  Before the node equivalence is defined, let $(n, i, n^\prime)$ be an outgoing edge of node $n$, and the set of all outgoing edges of node $n$ is denoted as $outgoing(n)$.  Two nodes $n_1$ and $n_2$ are \emph{equivalent}, denoted as $n_1 \equiv n_2$, if $n_1$ and $n_2$ are the same, or the following conditions are satisfied.
\begin{itemize}
\item $\forall_{(n_1, i, n^\prime_1)} \exists_{(n_2, i, n^\prime_2)}~n^\prime_1 \equiv n^\prime_2$.
\item $\forall_{(n_2, j, n^\prime_2)} \exists_{(n_1, j, n^\prime_1)}~n^\prime_1 \equiv n^\prime_2$.
\end{itemize}
Function \texttt{union} creates a new decision diagram for two input diagrams as follows.  Starting from the root nodes $n_1$ and $n_2$ of the two input decision diagrams, it creates a new node $n$ by following the rules shown below.
\begin{enumerate}
\item For each edge ${(n_1, i, n^\prime_1)} \in outgoing(n_1)$, if there is no ${(n_2, i, n^\prime_2)}$ in $outgoing(n_2)$, add $(n, i, n^\prime_1)$ to $outgoing(n)$.

\item For each edge ${(n_2, i, n^\prime_2)} \in outgoing(n_2)$, if there is no ${(n_1, i, n^\prime_1)}$ in $outgoing(n_1)$, add $(n, i, n^\prime_2)$ to $outgoing(n)$.

\item For each edge ${(n_1, i, n^\prime_1)} \in outgoing(n_1)$, if there is ${(n_2, i, n^\prime_2)}$ in $outgoing(n_2)$, add $(n, i, n^\prime)$ to $outgoing(n)$ where $n^\prime = \texttt{union}(n^\prime_1, n^\prime_2)$.
\end{enumerate} 
During the {union} operation, whenever a node is created, the unique table is checked to see if there is an existing node that is equivalent to the created one.  If there exists an equivalent node, such existing node is returned.  Otherwise, the created node is returned.   Multiple decision diagrams can exist at the same time, and they share the same unique node table.  This allows the node sharing among different decision diagrams.   

The {union} operation may need to make a large number of calls to function \texttt{union} on different pairs of nodes.  To avoid redundant work and improve efficiency, a cache is maintained to store pairs of nodes where function \texttt{union} has been applied and their corresponding results from the \texttt{union} operations.  If later a call to \texttt{union} on the same pair of nodes is made, then the cached result is returned instead of performing the expensive union operations again.  

After the {union} operation, if the input decision diagrams are no longer needed, their nodes can be removed by calling function \texttt{remove} on the root nodes $n$ as follows.
\begin{enumerate}
\item Decrement the reference count of $n$ by 1.

\item If the reference count of $n$ is $0$, remove $n$ from the unique node table, then for each edge ${(n, i, n^\prime)} \in outgoing(n)$, $\texttt{remove}(n^\prime)$. 
\end{enumerate}

More detailed description on multi-value decision diagrams and the commonly supported operations can be found in \cite{srinivasan1990}.  In our method, multi-value decision diagrams are used as a mechanism for storing the reachable states compactly, and the deletion operation for individual paths is not supported as it is not needed for the above purpose.  

\begin{figure}
\begin{center}
\begin{tabular}{cc}
\begin{minipage}{1in}
\begin{eqnarray*}
(0,0,0)\\
(0, 0, 1) \\
(0,1,0)\\
(0,1,1)\\
(1,1,1)\\
(1,1,2)\\
\end{eqnarray*}
\end{minipage}
&
\begin{minipage}{1.5in}
\begin{center}
\includegraphics[height=2in]{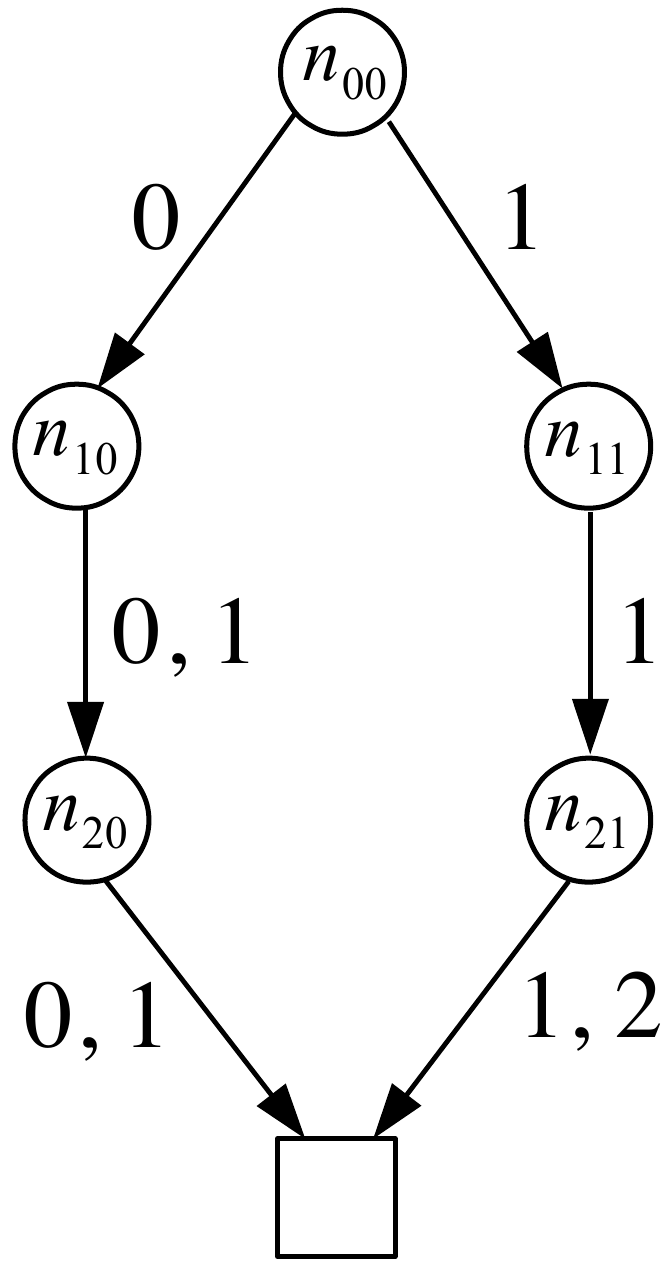}
\end{center}
\end{minipage} \\
(a) & (b)
\end{tabular}
\end{center}
\caption{\label{mdd-ex1} (a) A set of integer triples, (b) The corresponding MDD representation for the set shown in (a).}
\end{figure}

\begin{figure}
\begin{center}
\begin{tabular}{cc}
\includegraphics[height=2in]{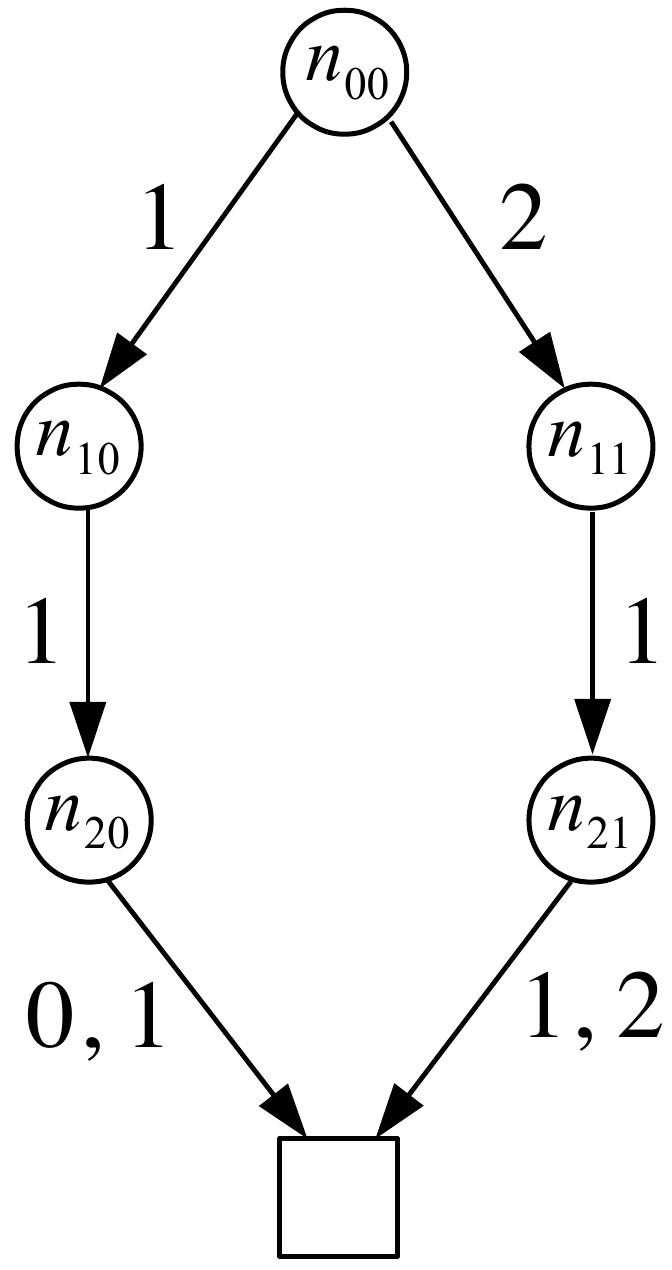} & \hspace{0.2in}
\includegraphics[height=2in]{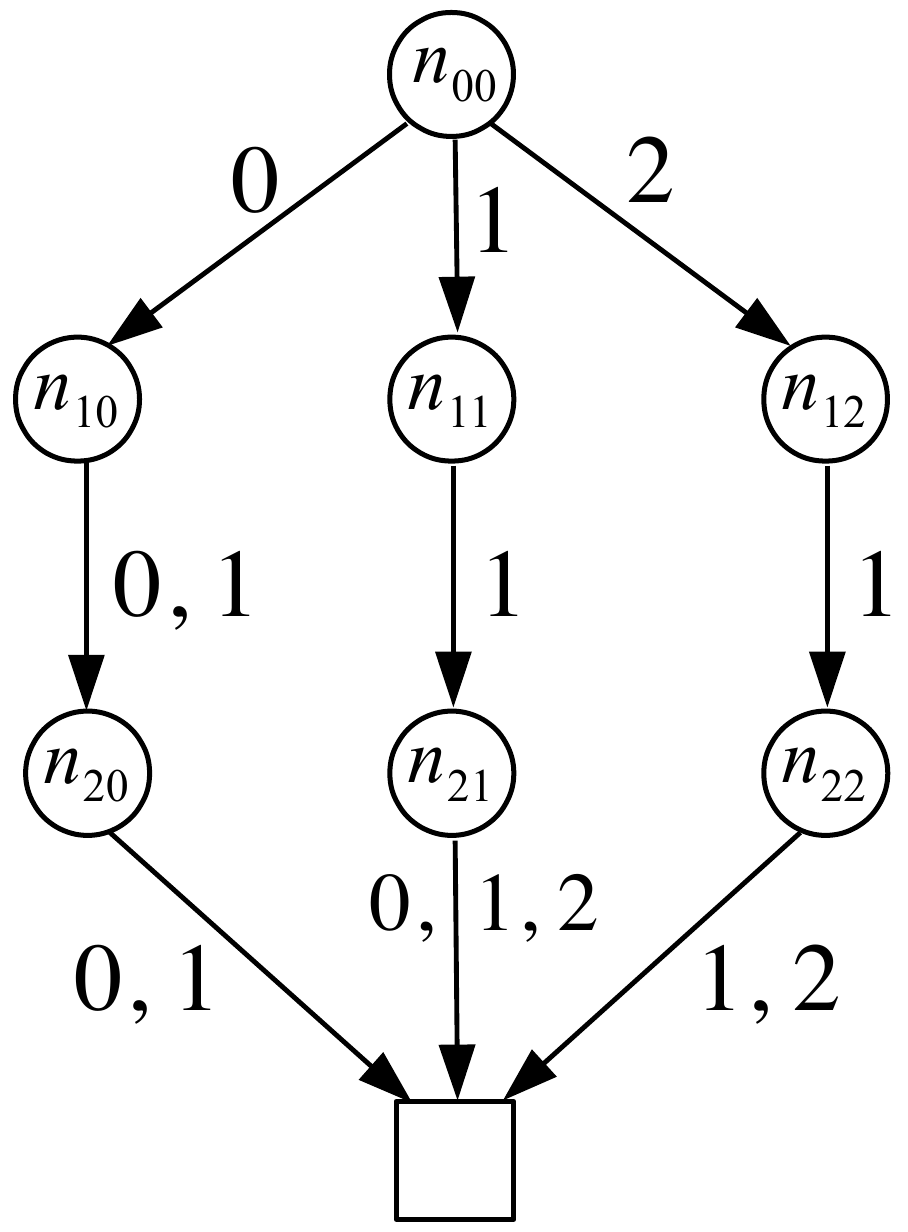} \\
(a) & \hspace{0.2in} (b)
\end{tabular}
\end{center}
\caption{\label{mdd-union-ex} (a) Another multi-value decision diagram, (b) The result from merging the MDD in Fig~.\ref{mdd-union-ex}(a) and the MDD in Fig.~\ref{mdd-ex1}(b) by the union operation.}
\end{figure}

The example in Fig.~\ref{mdd-ex1} shows the same set of integer triples in Fig~\ref{mdtree-ex1}(a) and the corresponding decision diagram.  Each edge is labeled with an integer denoting the index of some local state.  In the figure, an edge may be labeled with more than one integer.  In that case, it corresponds to multiple edges, each of which is labeled with an integer.  Another multi-value decision diagram is shown in Fig.~\ref{mdd-union-ex}(a), and the result from merging  it with the diagram shown in Fig.~\ref{mdd-ex1}(b) by function \texttt{union} is shown in Fig.~\ref{mdd-union-ex}(b).  It is straightforward to verify that all the paths in diagrams in Fig.~\ref{mdd-ex1}(b) and \ref{mdd-union-ex}(a) are included in the diagram shown in Fig.~\ref{mdd-union-ex}(b).  The resulting diagram represents a set of nine integer tuples.  Note that in the shown diagrams, the labels of the nodes are not used to distinguish nodes from different diagrams, therefore nodes from different diagrams with the same labels should not be viewed as the same nodes. 

\subsection{Using Decision Diagrams in Reachability Analysis}
\label{MDT}

A direct way to use the multi-value decision diagrams in the reachability analysis shown in Algorithm~\ref{basicSearch} is to replace the hash table for $stateTable$ with a decision diagram.  First, let $\texttt{mdd\_create}(\gSt)$ be a function that takes an integer tuple and returns a multi-value decision diagram with a single path corresponding to $\gSt$.  Then, the code in line~16 in Algorithm~\ref{basicSearch} can simply be replaced with the following line.
\centerline{$stateTable = \texttt{union}(stateTable,~\texttt{mdd\_create}(\gSt^\prime));$} 

Directly using the decision diagrams as shown above is simple, and can be very efficient in memory usage.  However, it is not very efficient in runtime.  This is due to a large number of \texttt{union} calls as one is made for every new state $\gSt^\prime$.  Since each \texttt{union} call can incur several operations on node creations and checks against the unique node table, a large number of \texttt{union} calls can lead to very high overhead when the number of reachable states is large.  

To address this problem, multi-value decision trees are used as buffers for a set of states before adding them together into the decision diagram $stateTable$.  The basic reason is that it is much more efficient to add a state into a decision tree.  On the other hand, a decision tree generally requires more memory to store the same number of states than a decision diagram.  In this method, a decision tree is used as a buffer.  The search algorithm adds states into the decision tree first.  When a sufficiently large number of states are added such that a preset memory threshold is exceeded, this decision tree is compressed to become a decision diagram, and consequently merged with the decision diagram pointed by $stateTable$ by function \texttt{union}.  This is a common idea of trading a reasonable amount of memory for higher runtime efficiency.  

Decision trees can be compressed into decision diagrams by function \texttt{compress}.  This function takes a decision tree rooted at node $n$, and performs the following operations recursively.
\begin{enumerate}
\item Create a new node $n_1$.

\item For each edge ${(n, i, n^\prime)} \in outgoing(n)$, 
	\begin{enumerate}
	\item $n^{\prime\prime} = \texttt{compress}(n^\prime)$.  
	\item Add edge ${(n_1, i, n^{\prime\prime})}$ into $outgoing(n_1)$.
	\end{enumerate}
	
\item Check if there exists an equivalent node to $n_1$ in the unique node table. If so, return the equivalent node; otherwise, return $n_1$.  
\end{enumerate}
Basically, function \texttt{compress} traverses a decision tree from the root node, and merges all equivalent nodes.

\section{Experiments}
\label{results}

\input{results-tables}

The multi-value decision trees and diagrams described in this paper are implemented in a package, and integrated into an asynchronous system verification tool $Platu$, an explicit model checker implemented in Java.  Experiments have been been performed on a number of examples.  These examples include asynchronous circuit designs from previously published papers \cite{Martin:FIFO,dill:PHD,stevens99relative,Yoneda96usingpartial,Myers:PhD}.  Other examples are selected from the BEEM benchmark suite for explicit model checking \cite{beem}.  This benchmark includes a large number of models of communication protocols, mutual exclusion algorithms, etc.  All the examples used in the experiments have medium or high complexity to show the difference in using the hash tables and the decision diagrams.  

In the experiments, all examples are ran by using the same depth-first search algorithm but with three different ways for storing reachable states: hash table, multi-value decision diagram, and multi-value decision diagram with a decision tree as buffer to reduce runtime overhead.  To have somewhat fair comparison, the model checker SPIN~\cite{holzmann:97} is not used in the experiments since SPIN is implemented in C while Platu is implemented in Java.  It is well known that the Java applications generally have serious memory overhead.  Instead, the state compression technique implemented in SPIN  is also implemented in Platu, and used as the base of state representation for all three methods representing reachable sets of states.  

Among all the examples, some have very large state space, and it can take enormous amount of time to find all reachable states.  The main purpose of using the decision diagrams for storing reachable states is to allow much larger number of states to be explored in a reasonable amount of time.  Hash tables can be accessed very efficiently, but usually exhaust memory also very quickly.  Therefore, for examples with small state space, hash tables are a better option.  On the other hand, for large examples, the decision diagrams would allow either the whole state space or a much larger portion of it to be stored with some extra runtime.  This allows larger examples to be handled, or improves verification coverage by exploring significantly more states.  In all experiments, upper bounds on time and memory are set to $900$ seconds and $2$~GB.  The results collected include the actual runtime, memory usage, and the total number of reachable states found at termination of the search algorithm.  In cases of time-out or memory-out, the number of reachable states is recorded.  All experiments are performed on a iMac desktop with a Intel Quad-core processor.  Only a single thread is used for all experiments.     
   
The results are shown in Table~\ref{table-1}.  The first column Column shows the different models. The numbers enclosed in parenthesis for the first few examples indicate the number of state variables.  Since these examples are asynchronous circuits, the type of their state variables is boolean.  The following examples are the models of mutual exclusion algorithms from the BEEM benchmarks \cite{beem}.  The remaining columns are divided into three groups, one for results from using each different state representation.  Columns under ``Base'' show results with the hash table is used.  Columns ``MDD'' show the results with the MDD alone used. Finally, the columns under ``MDD-Hybrid'' show the results with MDD used where a decision tree is used to speed up the search.  Among the five columns in each group, ``Time'' and ``Mem'' show the total runtime and the peak memory used during the search.  Note that the memory numbers include those used for storing reachable states and other necessary data structures for the search such as stacks.  Column ``$|S|$'' shows the total number of states found at the termination of the search.  Note that this number shows all reachable states of an example if the run does not time out ($<900$ second) and does not exhaust the allocated memory ($< 2000$ MB). 

The first conclusion that can be drawn from the table is that the search using a hash table is the fastest for almost all examples.  For examples with low complexity such as at.4 and fischer.3.  However, using a hash table also causes the search to exhaust 2~GB for large number of examples as shown by the cells in the table under ``Mem'' filled with 2000.  On the other hand, from the columns under ``MDD'', the search with MDD used is significantly slower, and it times out for most examples.  At the same time, using MDD allows more states to be searched with less memory.  Take dmeN5 as an example.  using MDD causes the search to time out, but over two times more states are found with slightly less memory than $2$ GB used.  The same conclusion is also observed for anderson.8, lamport.7, and so on.  For some other examples, the search with the MDD finds less number of states due to the overhead from converting states into the MDDs as explained before.    Finally, from the columns under ``MDD-Hybrid'',  the search in this mode finds much more states for almost every example within both limits of time and memory.  This shows the effectiveness of using decision trees as buffer for the MDD representation.  

To better compare these three representations, two more values are computed from the results for each example: search speed (SS) and state space density (SSD). The numbers in the column under SS show the numbers of states found per second.  This is a rough measure of how fast the search runs.  The numbers in the column under SSD show the numbers of states per MB of memory.  This is a rough measure of the efficiency of different representations for storing states.   For both measures, it is better and more efficient if the numbers are larger.  The averages of these two measurements are shown in the last row for every group.  From the average numbers, it can be seen alternatively that using the hash tables leads to the highest search speed but the least memory efficiency.  Using MDD alone leads to the highest memory efficiency, but it is the slowest to use.  The MDD with the decision tree buffering achieves good balance between runtime and memory as it is slight slower than using hash tables, and slightly less memory efficient than using the MDDs alone.  This good balance allows the search to find much more states for many examples.   Take szymamski.5 as an example, the search exceeds the memory limits by finding a little over 15 million states when the hash table is used, while the search runs out of time after finding  about 2 times more states with only 527 MB memory used when the MDDs alone are used.  However, the MDDs with buffering allow all reachable states to be found under both time and memory limits.   

From the experiments, for examples that take long time to run for the search using hash tables, using the MDDs would lead to less number of states to be found as  it is more expensive to use the MDDs even with faster buffering techniques.  On the other hand, for examples that have large state space, the MDDs are much more efficient.  Furthermore, buffering seems a very promising approach for handling larger designs.  Since the MDDs implemented in Platu are not intensively optimized, there might be more potential for making them even more efficient.  Meanwhile, it would be interesting to investigate if using the existing BDD packages such as CUDD \cite{cudd} could bring more efficiency as they commonly support variable recording, which is well known critical for efficiency and compactness of BDD representations.    

One last point to note is that in the work described in this paper, these different approaches are experimented to find out how states can be stored efficiently. The depth-first search algorithm itself remains the same.  This indicates that partial order reduction can be naturally integrated with the MDDs to allow much larger designs to be handled.  However, partial order reduction is not used in this paper. 

\section{Conclusion}
\label{conclusion}

This paper shows how the multi-value decision diagrams are used to store reachable states efficiently and compactly, which allows larger designs to be verified, or more states to be explored thus improving the verification coverage.  By using the decision trees as a buffering technique, the overhead of using the multi-value decision diagrams is significantly reduced.   In the future, it is interesting to find out how the binary decision diagrams compare against the multi-value decision diagrams in the same context.  Moreover, combining the decision diagrams and partial order reduction will also be investigated.



\input{hldvt2012.bbl}

\end{document}

%% file: results-tables.tex
\begin{sidewaystable*}[p]\small
\begin{center}
\caption{\label{table-1} Results from using different state representations on a set of asynchronous circuit designs and models of mutual exclusion examples.  Time is in seconds, and memory is in MBs. $|\stateset|$ is the numbers of states found at the end of reachability analysis.  Entries filled with $900$ under Time and $2000$ under Mem indicate time-out and memory.  Columns under ``Base'' show results using hash tables. Columns under ``MDD'' show results using MDDs, while those under ``MDD-hybrid'' show results using MDDs combined with decision trees as buffers.}

\begin{tabular}{|c||c|c|c|c|c||c|c|c|c|c||c|c|c|c|c|}
\hline
   & \multicolumn{5}{c||}{Base} & \multicolumn{5}{c||}{MDD} & \multicolumn{5}{c|}{MDD-Hybrid} \\
   \hline\hline

 Models & Time & Mem & $|S|$ & SS & SSD & Time & Mem & $|S|$ & SS & SSD & Time & Mem & $|S|$ & SS & SSD  \\ \hline
arbN5 (44) & 4.1 & 53 & 227472 & 55481 & 4292 & 5.3 & 32 & 227472 & 42919 & 7109 & 4.3 & 45 & 227472 & 52900 & 5055  \\ \hline
arbN7 (62) & 292 & 2000 & 12582972 & 43092 & 6291 & 548 & 1109 & 13801104 & 25184 & 12445 & 448 & 1639 & 13801104 & 30806 & 8420  \\ \hline
arbN9 (80) & 113 & 2000 & 9052417 & 80110 & 4526 & 320 & 2000 & 15892574 & 49664 & 7946 & 277 & 2000 & 18241258 & 65853 & 9121  \\ \hline
dmeN3 (33) & 2.77 & 49 & 267999 & 96751 & 5469 & 4.3 & 21 & 267999 & 62325 & 12762 & 2.4 & 29 & 267999 & 111666 & 9241  \\ \hline
dmeN4 (44) & 199 & 1725 & 15692028 & 78854 & 9097 & 322 & 155 & 15692028 & 48733 & 101239 & 152 & 560 & 15692028 & 103237 & 28021  \\ \hline
dmeN5 (55) & 149 & 2000 & 12590803 & 84502 & 6295 & 900 & 1881 & 38823343 & 43137 & 20640 & 900 & 2000 & 66746007 & 74162 & 33373  \\ \hline
fifoN8 (34) & 119 & 653 & 3572036 & 30017 & 5470 & 900 & 924 & 3268850 & 3632 & 3538 & 102 & 689 & 3572036 & 35020 & 5184  \\ \hline
fifoN10 (42) & 292 & 2000 & 9481673 & 32471 & 4741 & 900 & 301 & 334460 & 372 & 1111 & 767 & 2000 & 13164188 & 17163 & 6582  \\ \hline
mmu (55) & 900 & 1528 & 8284885 & 9205 & 5422 & 900 & 757 & 5492496 & 6103 & 7256 & 900 & 1701 & 9057024 & 10063 & 5325  \\ \hline
pipectrl (50) & 168 & 2000 & 10485516 & 62414 & 5243 & 376 & 2000 & 14444726 & 38417 & 7222 & 398 & 2000 & 22082395 & 55483 & 11041  \\ \hline
tagunit (48) & 269 & 2000 & 3302662 & 12278 & 1651 & 900 & 1919 & 2546805 & 2830 & 1327 & 900 & 1752 & 2668985 & 2966 & 1523  \\ \hline
anderson.6  & 141 & 2000 & 12013290 & 85201 & 6007 & 897 & 1724 & 18206917 & 20298 & 10561 & 348 & 1751 & 18206917 & 52319 & 10398  \\ \hline
anderson.8   & 41 & 2000 & 9022879 & 220070 & 4511 & 187 & 2000 & 12222460 & 65361 & 6111 & 192 & 2000 & 16141525 & 84070 & 8071  \\ \hline
at.4 &   102 & 1018 & 6597246 & 64679 & 6481 & 900 & 762 & 6159114 & 6843 & 8083 & 122 & 1624 & 6597246 & 54076 & 4062  \\ \hline
at.5 &   154 & 2000 & 10489013 & 68110 & 5245 & 900 & 1028 & 3651887 & 4058 & 3552 & 900 & 1752 & 12581219 & 13979 & 7181  \\ \hline
at.6 &   167 & 2000 & 10100451 & 60482 & 5050 & 900 & 1109 & 3651887 & 4058 & 3293 & 900 & 1752 & 11601860 & 12891 & 6622  \\ \hline
at.7 &   361 & 2000 & 12583081 & 34856 & 6292 & 900 & 506 & 3651887 & 4058 & 7217 & 900 & 1751 & 17686712 & 19652 & 10101  \\ \hline
bakery.8   & 214 & 2000 & 17042811 & 79639 & 8521 & 900 & 754 & 51820580 & 57578 & 68728 & 900 & 1751 & 76680860 & 85201 & 43793  \\ \hline
driving\_phils.3   & 1 & 13 & 436 & 436 & 34 & 1 & 14 & 436 & 436 & 31 & 1 & 14 & 436 & 436 & 31  \\ \hline
driving\_phils.4   & 204 & 2000 & 14039859 & 68823 & 7020 & 900 & 576 & 5217462 & 5797 & 9058 & 900 & 1752 & 28672594 & 31858 & 16366  \\ \hline
driving\_phils.5   & 78 & 2000 & 3599664 & 46150 & 1800 & 900 & 520 & 2268540 & 2521 & 4363 & 230 & 1750 & 3599663 & 15651 & 2057  \\ \hline
fischer.3 &   52 & 409 & 2896705 & 55706 & 7082 & 900 & 329 & 2363979 & 2627 & 7185 & 66 & 897 & 2896705 & 43889 & 3229  \\ \hline
fischer.4 &   25 & 342 & 1272254 & 50890 & 3720 & 471 & 461 & 1272254 & 2701 & 2760 & 47 & 1135 & 1272254 & 27069 & 1121  \\ \hline
fischer.5 &   340 & 2000 & 10780321 & 31707 & 5390 & 900 & 552 & 2505498 & 2784 & 4539 & 900 & 1752 & 5199544 & 5777 & 2968  \\ \hline
fischer.6 &   299 & 2000 & 8237316 & 27550 & 4119 & 900 & 533 & 2022330 & 2247 & 3794 & 900 & 1751 & 3811183 & 4235 & 2177  \\ \hline
fischer.7 &   365 & 2000 & 10633261 & 29132 & 5317 & 900 & 497 & 2717506 & 3019 & 5468 & 900 & 1752 & 7996790 & 8885 & 4564  \\ \hline
lamport.5 &   10 & 135 & 1066799 & 106680 & 7902 & 19 & 51 & 1066799 & 56147 & 20918 & 10.2 & 110 & 1066799 & 104588 & 9698  \\ \hline
lamport.6 &   1 & 11 & 519 & 519 & 47 & 1 & 14 & 519 & 519 & 37 & 1 & 12 & 519 & 519 & 43  \\ \hline
lamport.7 &   252 & 2000 & 18075514 & 71728 & 9038 & 900 & 310 & 33954095 & 37727 & 109529 & 497 & 1751 & 38717845 & 77903 & 22112  \\ \hline
lamport.8 &   1 & 11 & 50 & 50 & 5 & 1 & 11 & 50 & 50 & 5 & 1 & 11 & 50 & 50 & 5  \\ \hline
mcs.5 &   127 & 2000 & 12041513 & 94815 & 6021 & 900 & 1801 & 15106694 & 16785 & 8388 & 363 & 2000 & 23015234 & 63403 & 11508  \\ \hline
peterson.4   & 11 & 147 & 1119559 & 101778 & 7616 & 21 & 68 & 1119559 & 53312 & 16464 & 11 & 161 & 1119559 & 101778 & 6954  \\ \hline
peterson.5   & 109 & 2000 & 12583035 & 115441 & 6292 & 727 & 2000 & 23998663 & 33011 & 11999 & 517 & 2000 & 42200403 & 81626 & 21100  \\ \hline
peterson.6   & 88 & 2000 & 11110639 & 126257 & 5555 & 486 & 2000 & 13820828 & 28438 & 6910 & 309 & 2000 & 23897297 & 77338 & 11949  \\ \hline
peterson.7   & 222 & 2000 & 15975060 & 71960 & 7988 & 900 & 692 & 11957462 & 13286 & 17280 & 900 & 1752 & 64062366 & 71180 & 36565  \\ \hline
phils.6*  & 2.5 & 159 & 241660 & 96664 & 1520 & 14 & 141 & 241660 & 17261 & 1714 & 4 & 617 & 241660 & 60415 & 392  \\ \hline
phils.7   & 44 & 2000 & 6927036 & 157433 & 3464 & 900 & 1715 & 10253596 & 11393 & 5979 & 261 & 2000 & 10486177 & 40177 & 5243  \\ \hline
phils.8   & 7.2 & 437 & 914351 & 126993 & 2092 & 144 & 373 & 914351 & 6350 & 2451 & 13 & 1707 & 914351 & 70335 & 536  \\ \hline
szymanski.4   & 20.4 & 282 & 2313863 & 113425 & 8205 & 46 & 78 & 2313863 & 50301 & 29665 & 21 & 204 & 2313863 & 110184 & 11342  \\ \hline
szymanski.5   & 155 & 2000 & 15908655 & 102636 & 7954 & 900 & 527 & 48526313 & 53918 & 92080 & 867 & 1751 & 79514858 & 91713 & 45411  \\ \hline
Average &  &  &    & 69875 & 5220 &  &  &  & 22155 & 16269 &  &  &  & 49263 & 10712  \\ \hline
\end{tabular}
\end{center}
\end{sidewaystable*}